\documentclass[3p,times,procedia]{elsarticle}
\usepackage{nupha_ecrc}
\usepackage{lineno}
\usepackage[margin=2mm]{caption}

\volume{00}

\firstpage{1}

\journalname{Nuclear Physics A}

\runauth{}

\jid{nupha}

\jnltitlelogo{Nuclear Physics A}

\usepackage{amssymb}

\usepackage[figuresright]{rotating}

\newcommand*\snn{\sqrt{s_{NN}}}

\newcommand*\tpt{t\approx p^2_{\mathrm{T}}}
\newcommand*\pt{p_{\mathrm{T}}}

\newcommand*\mpmm{\mu^+\mu^-}
\newcommand*\epem{\mathrm{e}^+\mathrm{e}^-}
\newcommand*\Mee{M_{\mathrm{ee}}}

\begin{document}

\begin{frontmatter}



\dochead{}

\title{Dilepton Production in p$+$p, Au$+$Au collisions at $\sqrt{s_{NN}}$~=~200~GeV and U$+$U collisions at $\sqrt{s_{NN}}$~=~193~GeV}


\author{James D. Brandenburg (for the STAR Collaboration)}

\address{Rice University, Houston, TX, USA}

\begin{abstract}
In this contribution we report $e^{+}e^{-}$ spectra with various invariant mass and p$_{T}$ differentials in Au$+$Au collisions at $\sqrt{s_{NN}}$=200 GeV and U$+$U collisions at $\sqrt{s_{NN}}$=193 GeV. The structure of the t ($\tpt$) distributions of these mass regions will be shown and compared with the same distributions in ultra-peripheral collisions. Additionally, this contribution discusses first measurements of $\mu^{+}\mu^{-}$ invariant mass spectra from STAR's recently installed Muon Telescope Detector (MTD) in p$+$p and Au$+$Au collisions at $\sqrt{s_{NN}}$ = 200 GeV.
\end{abstract}


\end{frontmatter}

\section{Introduction}
\label{sec:intro}
Heavy-ion collisions like those produced at the Relativistic Heavy Ion Collider (RHIC) produce a hot and dense strongly interacting medium of quarks and gluons\cite{QGP_0,QGP_1,QGP_2,QGP_3}. Dileptons ($l^{+}l^{-}$) are produced throughout all stages of the medium's evolution and escape with minimum interaction with the strongly interacting medium. For this reason, $l^{+}l^{-}$ pair measurements play an essential role in the study of the hot and dense nuclear matter created in heavy-ion collisions. Dileptons in the low invariant mass region (up to M$_{ll}\sim$1~GeV/c$^2$) retain information about the in-medium modification of vector mesons while dileptons in the intermediate mass region (extending out to M$_{ll}\sim$3 GeV/c$^2$) predominantly originate from charm decays and thermal radiation of the medium. At higher invariant masses ( above M$_{ll}\sim$3 GeV/c$^2$ ) dileptons originate from the Drell-Yan process and from the decay of heavy quarkonia such as $J/\psi$ and $\Upsilon$.  \par
Recent studies of the $J/\psi$ yields in peripheral Pb$+$Pb collisions at $\snn$ = 2.7 TeV by the ALICE collaboration showed significant excess at very low momentum transfers ($\pt$ $<$ 0.3~GeV/c)\cite{alice}. STAR has also measured significant excess $J/\psi$ yields in peripheral Au+Au collisions at $\snn$ = 200 GeV and U+U collisions at $\snn$ = 193 GeV for $\pt$ $<$ 0.15 GeV/c\cite{wangmei}.
These observations cannot be explained by hadronic production mechanisms and may be evidence of dilepton production from photon-photon or photon-nucleus interactions. These types of interactions are studied in great detail in ultra-peripheral collisions (UPC) where the impact parameter, b, is larger than two times the nuclear radius and no hadronic interaction takes place. This motivated STAR to measure the $e^{+}e^{-}$ pair production in a wider invariant mass region (M$_{ee}~<~$4~GeV/c$^2$) at very low p$_T$ to further investigate the possible production mechanisms responsible for the observed $J/\psi$ excess at low $\pt$ in peripheral A$+$A collisions.

\section{Low $\pt$ $\epem$ Analysis and Results}
\label{sec:results}

\begin{figure}
	\centering
	\begin{minipage}[t]{0.62\textwidth}
		\includegraphics[width=0.99\textwidth]{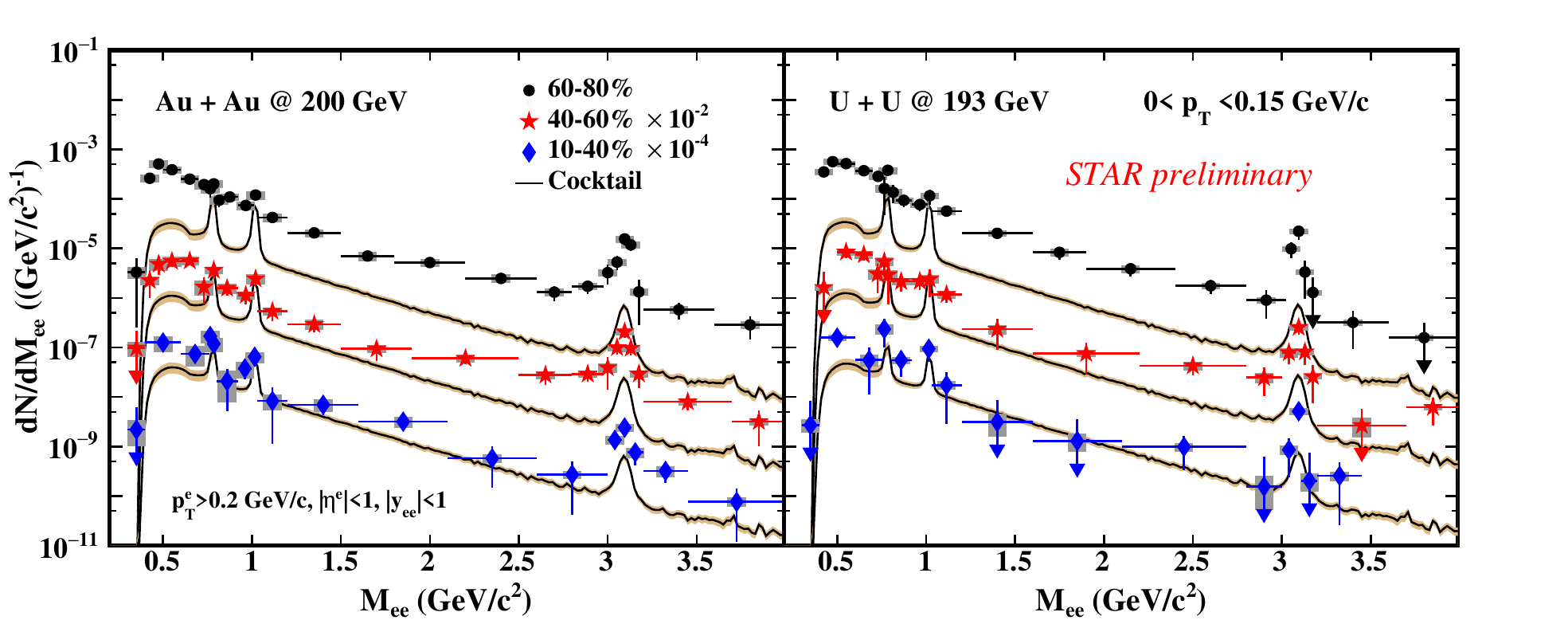} 
		
		\caption{\label{fig:low_pt_inv_mass} (color online) The low $\pt$ (0$<$$\pt$$<$0.15 GeV/c) $\epem$ invariant mass distribution for Au+Au collisions at $\snn$=200 GeV (left) and U+U collisions at $\snn$=193 GeV(right). The invariant mass distributions for collisions with 60-80\%, 40-60\%, and 10-40\% centrality are shown. For each centrality the corresponding hadronic cocktail contribution is shown as a solid black line with the uncertainty as a shaded region. Statistical and systematic uncertainties on the data points are indicated with vertical lines and shaded boxes, respectively.} 
	\end{minipage} 
	\begin{minipage}[t]{0.37\textwidth}
		\includegraphics[width=0.9\textwidth]{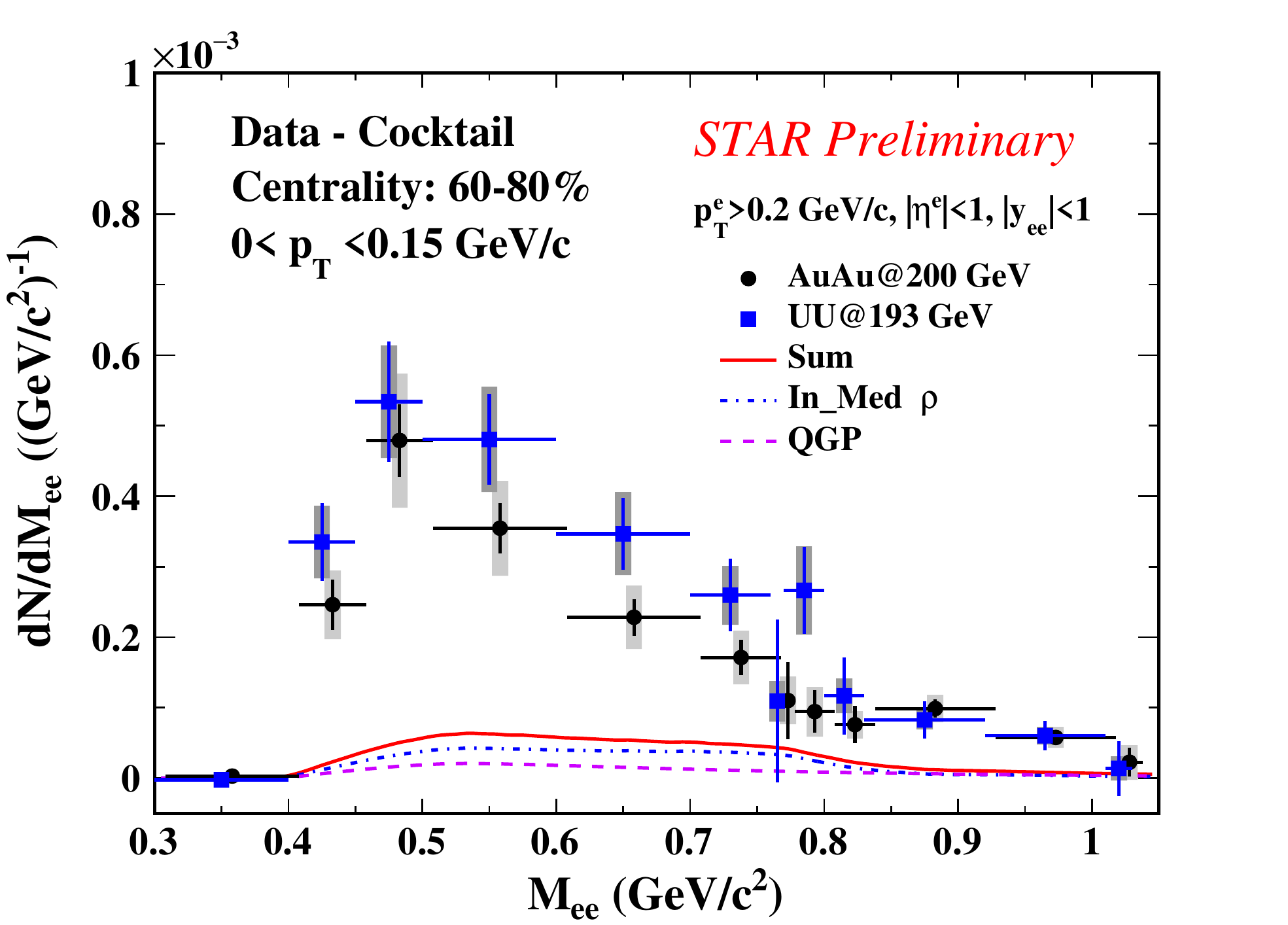} 
		
		\caption{ \label{fig:low_pt_excess} (color online) The low $\pt$ $\epem$ excess yield with respect to the hadronic cocktail for Au$+$Au and U$+$U collisions with 60-80\% centrality. The additional contributions expected from in-medium modified $\rho$ production, QGP thermal radiation, and the sum of the two are shown. } 
	\end{minipage}
	
\end{figure}

STAR's Time Projection Chamber (TPC) and Time-of-Flight Detector (TOF) are used in the measurement of the $\epem$ invariant mass spectra reported in this contribution\cite{STAR, STAR_TPC, STAR_TOF}. The TPC provides charged particle tracking, momentum measurement, and particle identification through ionization energy loss measurement. The TOF detector provides precise particle velocity ($\beta$) measurements. The particle identification power of these two detectors is combined to provide clean electron identification out to $\pt\sim$3 GeV/c. \par
Figure \ref{fig:low_pt_inv_mass} shows the $\epem$ invariant mass distribution from Au$+$Au collisions at $\snn$=200 GeV and U$+$U collisions at $\snn$=193 GeV for $\pt$ $<$ 0.15 GeV/c. The $\epem$ invariant mass distribution is shown for 60-80\%, 40-60\%, and 10-40\% collision centralities. For each case the expected contribution from hadronic sources (excluding the $\rho$ meson) is shown as a solid black line. In 60-80\% central Au$+$Au and U$+$U collisions there is a significant excess visible with respect to the corresponding hadronic cocktail. The excess is less significant in 40-60\% central collisions while the data from 10-40\% central collisions is consistent with the expectation from the hadronic cocktail.\par
The excess yield with respect to the hadronic cocktail is shown in Fig. \ref{fig:low_pt_excess} for 60-80\% central Au$+$Au and U$+$U collisions along with the additional contributions expected from an in-medium modified $\rho$ and thermal radiation from the Quark Gluon Plasma (QGP). The sum of these additional contributions is insufficient to explain the observed excess. In this contribution we will investigate the possibility that the additional dilepton yield observed at low $\pt$ results from UPC-like photon-photon or photon-nucleus interactions.\par

\begin{figure}[!ht]
	\centering 
	\begin{minipage}[t]{0.66\textwidth}
	\includegraphics[width=0.99\textwidth]{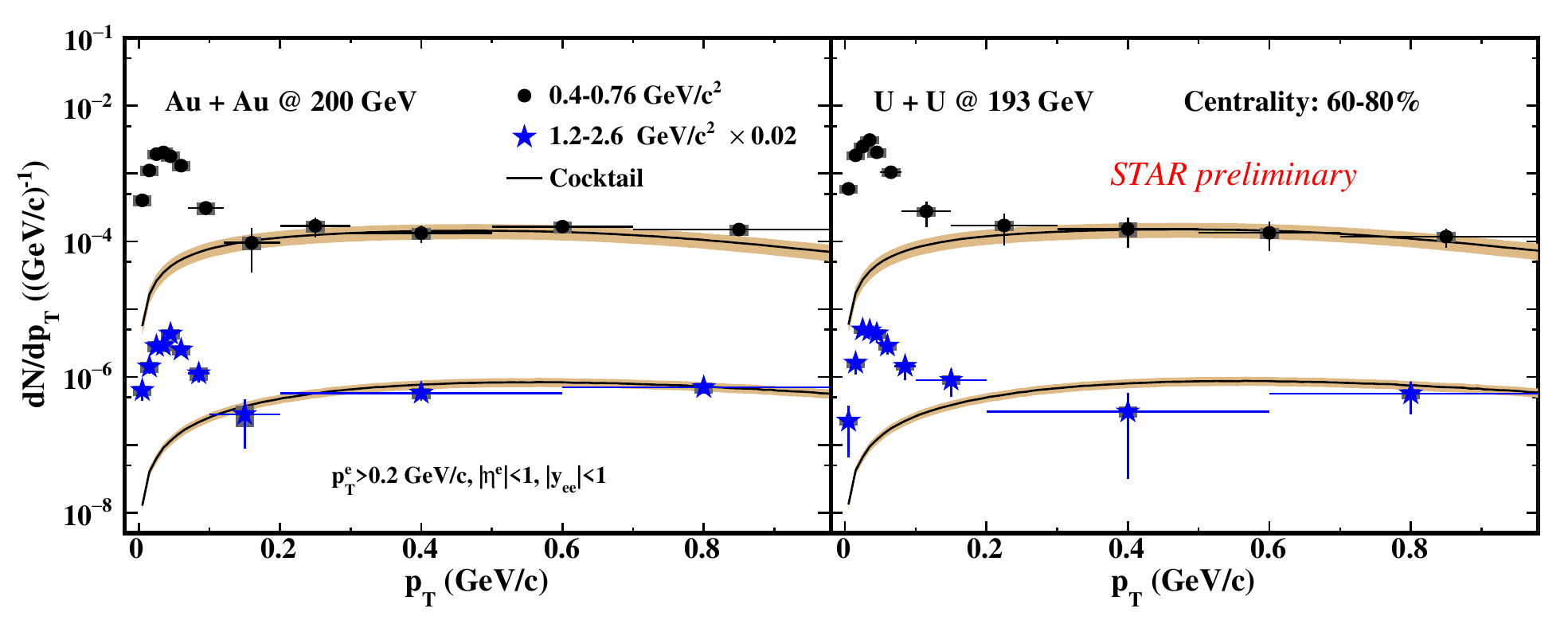} 
	
	\caption{ \label{fig:low_pt_spectra} (color online) The $\pt$ distribution of $\epem$ pairs in the $\rho$-like mass region (0.4 $<$ $\Mee$ $<$ 0.76 GeV/c$^2$) and the continuum region (1.2 $<$ $\Mee$ $<$ 2.6 GeV/c$^2$) shown for Au+Au collisions at $\snn$~=~200 GeV (left) and U+U collisions at $\snn$~=~193 GeV(right). For each case the expected contribution from the hadronic cocktail sources is shown in a solid black line with the uncertainty shown as a shaded band. } 
	\end{minipage}
	\begin{minipage}[t]{0.32\textwidth} 
		\includegraphics[width=0.9\textwidth]{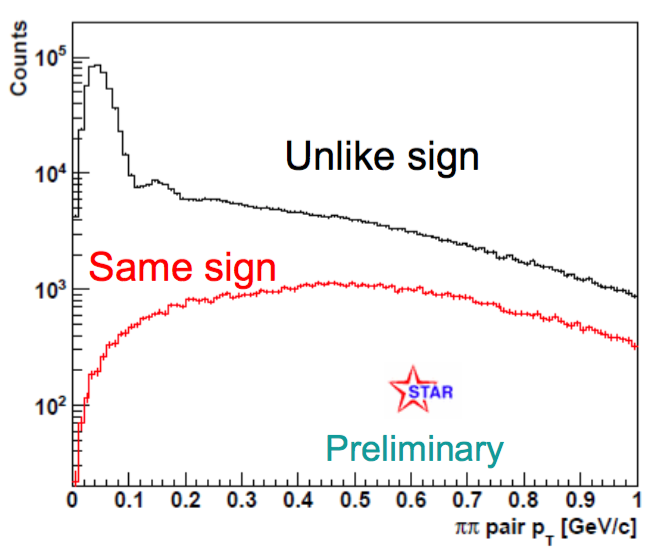} 
		
		\caption{ \label{fig:upc_rho0_pt} (color online) The $\pt$ distribution of coherently produced $\rho^0\rightarrow\pi\pi$ from STAR UPC events in Au$+$Au at $\snn$=200 GeV. } 
	\end{minipage}
\end{figure}

The characteristics of the excess dilepton production can be further studied by investigating the $\pt$ distribution in various invariant mass regions. In Fig. \ref{fig:low_pt_spectra} the $\pt$ distribution is shown for the $\rho$-like mass region (0.4~$<$~$\Mee$~$<$~0.76~GeV/c$^2$) and the continuum region (1.2~$<\Mee<$~2.6~GeV/c$^2$) in 60-80\% central Au$+$Au and U$+$U collisions. The hadronic cocktail adequately describes the data for $\pt>$~0.15~GeV/c in both collision species and both mass regions. The entire observed excess with respect to the hadronic cocktail is found below $\pt\approx$~0.15~GeV/c. The structure of the $\pt$ distribution in Fig. \ref{fig:low_pt_spectra} appears similar to the $\pt$ distribution for coherent photoproduction of $\rho^0$ in UPCs seen in Fig. \ref{fig:upc_rho0_pt}. This similarity is motivation to investigate whether or not coherent photoproduction could explain the observed excess production of low-$\pt$ dielectron pairs in peripheral A$+$A collisions. Photon-nucleus interactions are characterized by strong coupling ($Z\alpha_{\mathrm{em}}$$\sim$0.6) and a photon wavelength $\lambda$~=~$h/p$~$>$~$R_{\mathrm{A}}$, where $R_{\mathrm{A}}$ is the radius of the colliding nuclei\cite{UPC_one}. This relationship between the photon wavelength and $R_{\mathrm{A}}$ limits the photon's transverse momentum such that the photon's $\pt$ is less than $h/R_{\mathrm{A}}$~($\pt$$<$$\sim$30 MeV for heavy ions). The strong coupling strength and low $\pt$ limit of the photoproduction mechanism leads to a large production cross-section at low $\pt$ which falls off rapidly at higher momentum transfers. \par 

Additionally, the coherent photon-nucleus interaction is the result of two indistinguishable processes in which the photon can be emitted from A$_1$ or A$_2$, i.e. from either nucleus. For exclusive vector meson production the amplitude of these two processes have opposite signs which results in strong destructive interference patterns in the $\pt$ distribution\cite{UPC_inf}. One experimental signature of this is a strong downturn in the $\tpt$ distribution at very low values of $|t|$ followed by an exponential decrease in the yield as $|t|$ increases. Then, at higher values of $|t|$ the total yield flattens out as the other production mechanisms dominate over the coherent photoproduction mechanism. Figure \ref{fig:low_pt_t_dist} shows the $t$ distribution for peripheral Au$+$Au and U$+$U collisions for the $\rho$-like and continuum mass regions. These $t$ distributions are fit to $A e^{-B t}$ in the low $t$ region excluding the data point with lowest $t$ value. The large slope parameters from the $t$ distribution fits are similar to the slope parameters found in coherent photoproduction of $\rho^0$ in UPCs\cite{UPC_two}. However, more data points with smaller uncertainties are needed to make a definitive conclusion as to whether or not coherent photoproduction can explain the observed excess dilepton yield at low $\pt$ in peripheral A$+$A collisions.

\section{ $\mpmm$ Measurements with the MTD at STAR}
\label{sec:muons}
Installation of the Muon Telescope Detector (MTD) at STAR was completed in 2014. The MTD, located outside the STAR magnet steel, provides $\sim$45\% azimuthal coverage in $-$0.5$<$$\eta$$<$0.5\cite{MTD_calib}. The precise timing ($\sigma$$\sim$100~ps) and spatial resolution ($\sigma$$\sim$1~cm) of the MTD aids in the identification of muons\cite{MTD_pid}. With the addition of the MTD, STAR has collected data with a dedicated dimuon trigger in p$+$p and Au$+$Au collisions at $\snn$=200 GeV. These datasets allow the $\mpmm$ invariant mass spectra to be measured over a large mass range for the first time at STAR. Measurements of the $\mpmm$ invariant mass spectra will allow STAR to conduct new studies of the in-medium modification of the $\rho$ meson in the LMR where the $\mpmm$ channel suffers from fewer background sources compared to the $\epem$ channel. Measurement of the IMR in the $\mpmm$ channel will also allow for new studies of the QGP thermal radiation. The new muon identification capabilities at STAR also allow for $e - \mu$ correlation studies which may help reduce the uncertainty on the significant contribution from semi-leptonic charm decays in the IMR\cite{MTD_emu}. Dedicated dimuon-triggered datasets were collected from Au$+$Au collisions at $\snn$=200 GeV in 2014 and 2016. Preliminary analysis of these datasets show S/B of $\sim$1/10 in 60-80\% central collisions. This S/B ratio is significantly higher than the S/B ratio of $\sim$1/100 observed in $\epem$. \par
\begin{figure}
	\centering 
	\begin{minipage}[t]{0.45\textwidth}
	\includegraphics[width=1.0\textwidth]{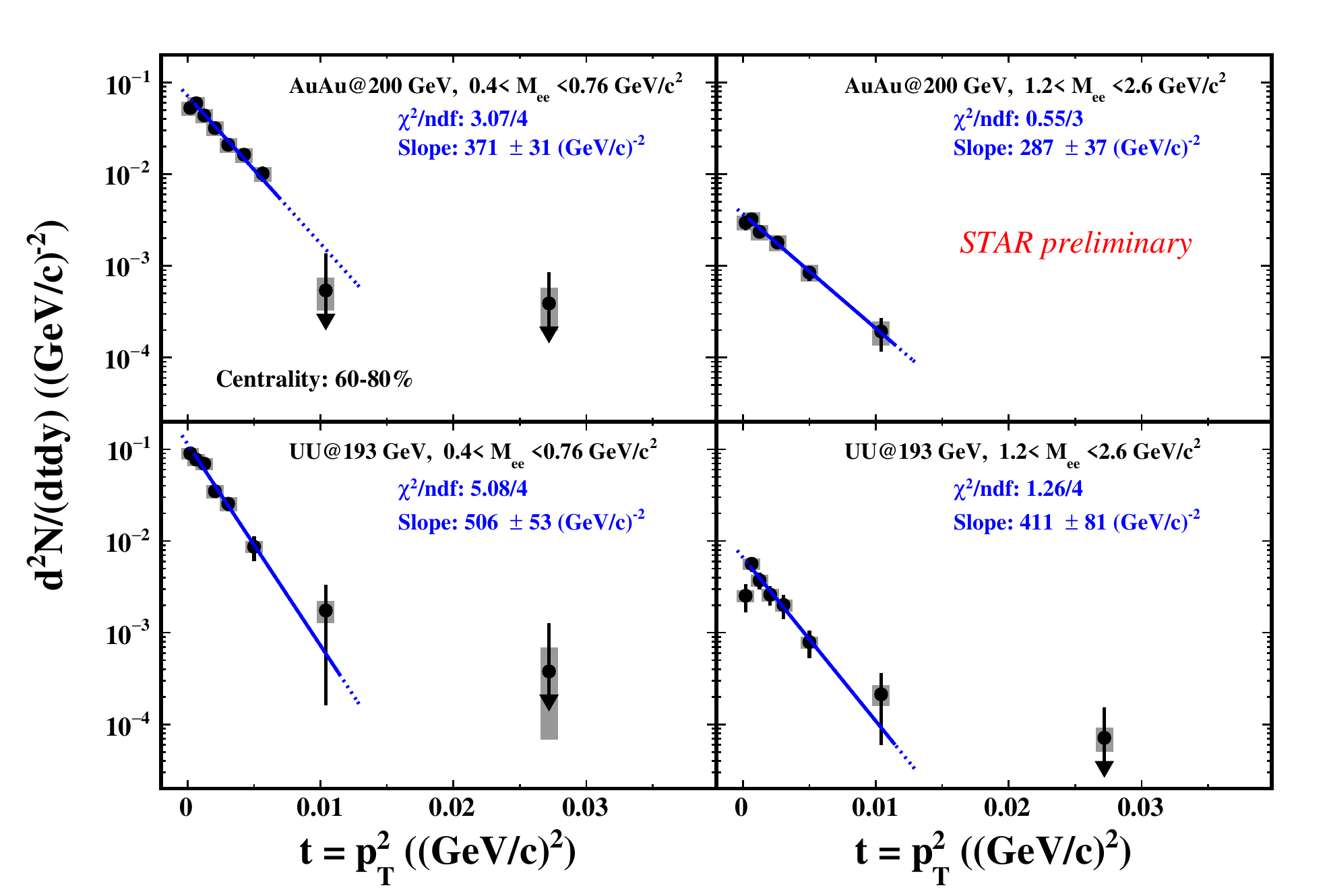} 
	
	\caption{ \label{fig:low_pt_t_dist} (color online) The $\tpt$ distributions for $\epem$ pairs in two different mass regions shown for 60-80\% central Au+Au and U$+$U collisions. } 
	\end{minipage}
	\begin{minipage}[t]{0.45\textwidth}
		\includegraphics[width=1.0\textwidth]{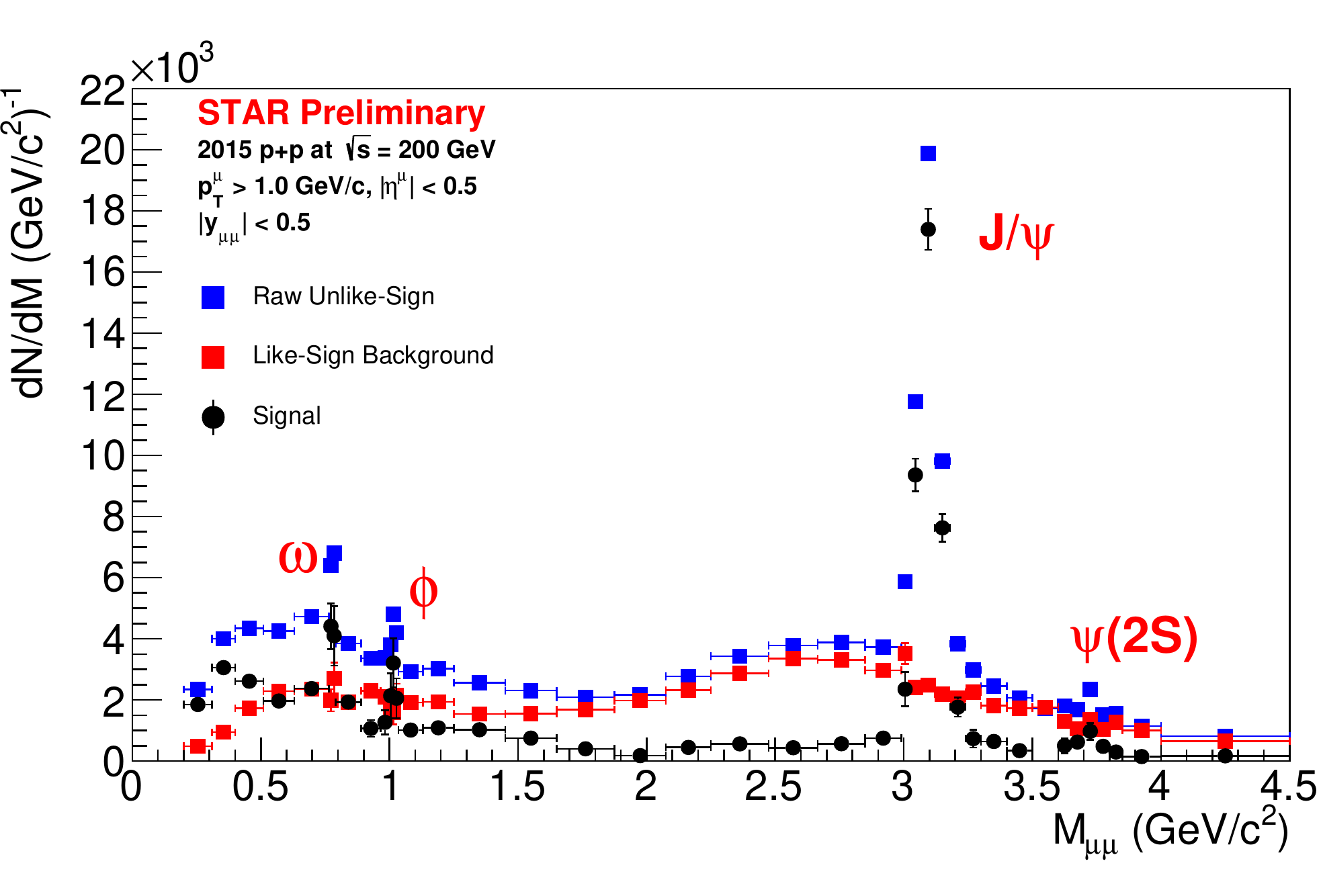} 
		
		\caption{ \label{fig:dimuon_inv_mass} (color online) The $\mpmm$ invariant mass distribution for p$+$p collisions at $\sqrt{s}$=200 GeV. }
	\end{minipage}

\end{figure}

The p$+$p data collected in 2015 will provide a high quality baseline with which to compare future STAR measurements of the $\mpmm$ invariant mass spectra in central Au$+$Au collisions. Figure \ref{fig:dimuon_inv_mass} shows the raw $\mpmm$ distribution in data from 2015 p$+$p collisions at $\sqrt{s}$=200 GeV. Also shown is the like-sign ($(\mu^{+}\mu^{+} + \mu^{-}\mu^{-})/2$) distribution and the difference between the unlike-sign and like-sign distributions (signal). The $\mpmm$ signal shows clear peaks corresponding to the $\omega$, $\phi$, $J/\psi$, and $\psi(2S)$. The full analysis of the 2015 p$+$p and 2014 Au$+$Au datasets is ongoing.

\section{Summary}
\label{sec:summary}

In this contribution we have reported on recent STAR measurements of the $\epem$ invariant mass spectra at low $\pt$ in peripheral A$+$A collisions. If the observed excess dilepton production is the result of coherent photoproduction of vector mesons in peripheral A$+$A collisions then this may provide a novel probe for studying the QGP medium. For instance, if a vector meson is produced through photoproduction with very low $\pt$ followed by a hadronic interaction then the vector meson may remain in the vicinity of the strongly interacting medium until it has decayed into a dilepton pair. In this case the dileptons from the photo-produced vector meson could be used as a probe of the medium characteristics. However, more data with smaller uncertainties will be needed to support or reject the hypothesis that coherent photoproduction is responsible for the observed dilepton excess in peripheral A$+$A collisions at very low $\pt$. STAR plans to collect data from $^{96}_{44}$Ru$+$$^{96}_{44}$Ru and $^{96}_{40}$Zr$+$$^{96}_{40}$Zr collisions in 2018. Data from these A=96 isobar collisions will help determine the exact production mechanism responsible for the excess since the photon flux from photon-photon and photon-nucleus interactions scale $\propto Z^4$ and $\propto Z^2$, respectively. \par
Finally, the full installation of the MTD at STAR in 2014 has allowed STAR to collect dedicated dimuon-triggered datasets in p$+$p and Au$+$Au collisions at $\snn$=200 GeV. These datasets will allow new studies of the $\mpmm$ invariant mass spectra over a large mass range for the first time at STAR. The 2015 p$+$p dataset provides a high quality baseline for STAR measurements in central Au$+$Au collisions. The Au$+$Au datasets collected in 2014 and 2016 will provide new opportunities to study the in-medium broadening of the $\rho$ meson in the LMR and to investigate the contributions from QGP thermal radiation and semi-leptonic charm decays in the IMR.





\bibliographystyle{elsarticle-num}
\bibliography{<your-bib-database>}



\end{document}